# Localized Feature Aggregation Module for Semantic Segmentation

Ryouichi Furukawa , *Kazuhiro Hotta*

*Meijo University, Japan*

*Abstract*— **We propose a new information aggregation method which called "Localized Feature Aggregation Module" based on the similarity between the feature maps of an encoder and a decoder. The proposed method recovers positional information by emphasizing the similarity between decoder's feature maps with superior semantic information and encoder's feature maps with superior positional information. The proposed method can learn positional information more efficiently than conventional concatenation in the U-net and attention U-net. Additionally, the proposed method also uses localized attention range to reduce the computational cost. Two innovations contributed to improve the segmentation accuracy with lower computational cost. By experiments on the Drosophila cell image dataset and COVID-19 image dataset, we confirmed that our method outperformed conventional methods.**

## I. Introduction

Semantic segmentation requires to learn both semantic information and spatial information to improve the accuracy. Therefore, various segmentation methods for learning spatial information have been studied [1,2,17]. For example, U-net [3] conveys encoder's information to the decoder through skip connections and teaches positional information to the decoder by concatenation. Attention U-Net [18,19] also conveys encoder's information to the decoder through skip connections and teaches positional information by creating a mask for positional information from the feature maps of the encoder and decoder. The method multiplied the mask by the feature maps of the decoder. However, those two methods did not consider the direct relationship between the feature maps of the encoder and decoder. Thus, they do not seem to be the best way to teach position information from the encoder.

We propose localized feature aggregation module based on source-target-attention [5,9] to compute the relationship between feature maps in the encoder and decoder. Localized feature aggregation module is shown in Fig. 1. The module calculates the similarity between encoder's feature maps with superior positional information and the decoder's feature maps with superior semantic information. The similarity is measured by the inner product between the encoder's and decoder's features in source-target-attention. The model with the proposed module learns to emphasize the pixels with high similarity between the encoder and decoder while suppressing the pixels with low similarity. Therefore, the proposed method can learn location information efficiently. However, the computational cost of source-target-attention in our module is too high for using GPU, and we must reduce the computational cost. In order to reduce the computational cost, we employed the attention mechanism locally as shown in Fig. 1. As a result, the proposed module is able to incorporate more spatial information, and it is expected to improve the accuracy further.

In experiments on the Drosophila cell image dataset [14] and COVID-19 image dataset [28], we compared our method with conventional U-net and attention U-Net. We found that our method improved the accuracy by 1.46% compared to U-net and 0.829% compared to attention U-net on the Drosophila cell image dataset. The accuracy of our method on COVID-19 image dataset was 2.60% higher than U-net and 2.45% higher than attention U-net. In addition, the effectiveness of our method was also confirmed by ablation studies.

This paper is organized as follows. Section 2 describes the related works. The details of the proposed is presented in section 3. Section 4 shows the experimental results on two datasets. Finally, conclusion and future works are described in section 5.

## II. Related works

### A. Semantic segmentation

The basic structure for semantic segmentation [10-13] is the encoder-decoder structure. U-Net is a famous encoder-decoder model. The most important characteristics of U-Net is the skip connection between encoder and decoder. By using skip connections, encoder's feature maps and the decoder's feature maps are concatenated to convey spatial information. Therefore, the spatial information is complemented, and each pixel can be more accurately assigned to the class label. Attention U-net also conveys encoder's information to the decoder through skip connections. The most important characteristics of attention U-Net is to create a mask for positional information from the feature maps in the encoder and decoder. Attention U-net can suppress irrelevant regions in an image and highlight salient features that are useful for a specific task by the mask. Therefore, it can generate better feature maps than U-net, and each pixel can be more accurately classified.

However, those two methods did not consider the direct relationship between the feature maps in the encoder and decoder. We are able to create better feature maps by learning the relationships between pixels in the feature maps of encoders with superior positional information and those of decoders with superior semantic information. In this paper, to use the relationship between the feature maps of encoder and decoder effectively, we used source-target-attention [5] between feature maps instead of concatenation.

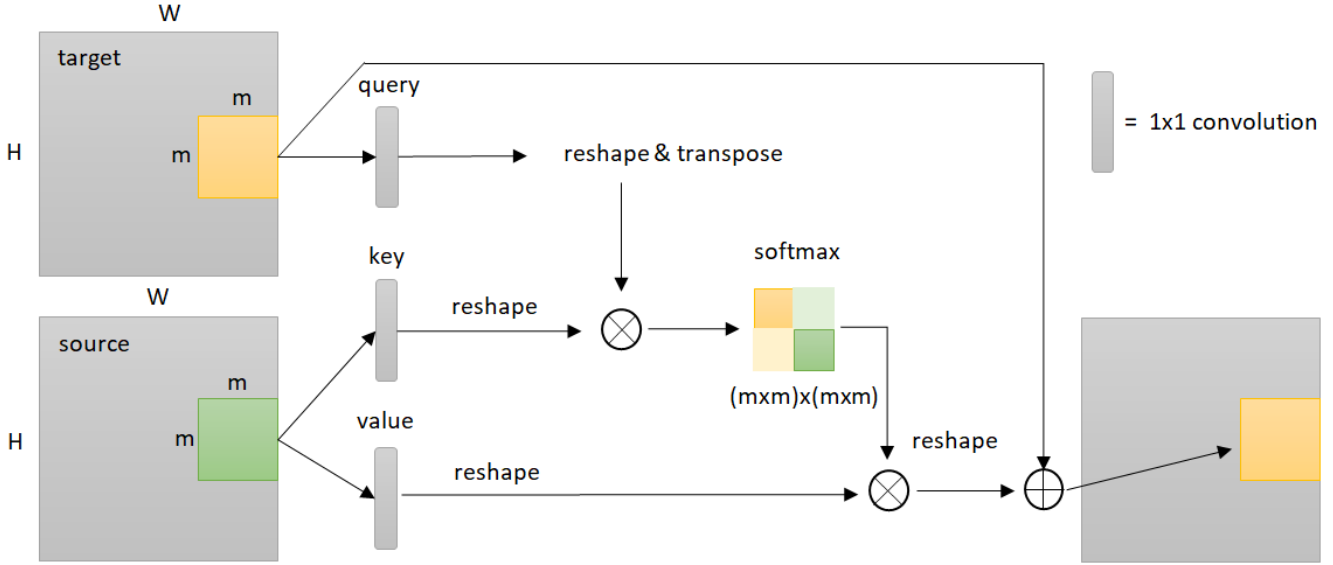

Fig. 1: Localized feature aggregation module

*B. Attention mechanism*

Attention mechanism has been used for various tasks such as machine translation [4,5], language processing (NLP) for image captioning [20], image classification [21,22] and segmentation [23-25]. One of the successful method is SENet [29] which emphasizes important channels. There is also a derived model of SENet that efficiently builds interdependencies in the spatial direction [33,34]. However, SEnet is not suitable for aggregating the information between different features.

Self-attention and source-target-attention are the methods used for machine translation [5]. Self-attention aggregates the features by the similarity between the same feature maps. Source-target-attention aggregates the features by the similarity between different feature maps. Self-attention networks [30] are very popular due to their capability of building spatial or channel attention. Typical examples are NLNet [30], GCNet [31], SCNet [32] and CCNet [8], and all of them exploited non-local attention mechanisms to capture contextual information.

In this paper, we used source-target-attention because we would like to aggregate the information that have the similarity between different features. We aggregate the features using the similarity between feature maps in the encoder and decoder. We show the effectiveness of our method in comparison with conventional concatenation by experiments.

III. PROPOSED METHOD

*A. Feature aggregation module*

We propose a new feature aggregation method that computes the similarity between the feature maps in an encoder and a decoder. The details of feature aggregation module is shown in Fig.1. Feature aggregation module consists mainly of the source-target-attention. Source-target attention used in machine translation is a mechanism that predicts what the next piece of information (Key, Value) will be when information (Query) is received at a certain time. In other words, the source-target-attention ascertains which information in the Value is to be obtained based on the information in the Query by the inner product with the Key. Since segmentation must classify each pixel in an image, it is necessary to recover the location information from the feature maps aggregated by the encoder. We defined the feature maps in encoder as the Query which is superior in positional information. The feature maps in decoder which is superior in semantic information are used as the Key and Value. Feature aggregation module receives the information of the encoder with superior spatial information and calculates the similarity with the feature maps in the decoder with superior semantic information, and we obtain better feature maps with the positional and semantic information.

The mechanism of feature aggregation module is explained. P is defined as the feature map of the encoder, and Q is defined as the feature map of the decoder. Pixel p is defined as a pixel on the encoder's feature map, and pixel q is defined as a pixel on the decoder's feature map. The functions to be linearly transformed into query, key, and value are $f_q(\cdot)$, $f_k(\cdot)$, and $f_v(\cdot)$. $W_{pq}$ is the attention map based on the similarity between feature maps P and Q.

$$W_{pq} = softmax_q(f_q(p)^T f_k(q)) \quad (1)$$

$Y_p$ indicates the result of aggregating the similarity of each pixel in feature map $Q$ to pixel $p$ belonging to feature map $P$ using the attention map.

$$Y_p = \sum_{q \in H2xW2} W_{pq} \cdot f_v(Q_q) \quad (2)$$

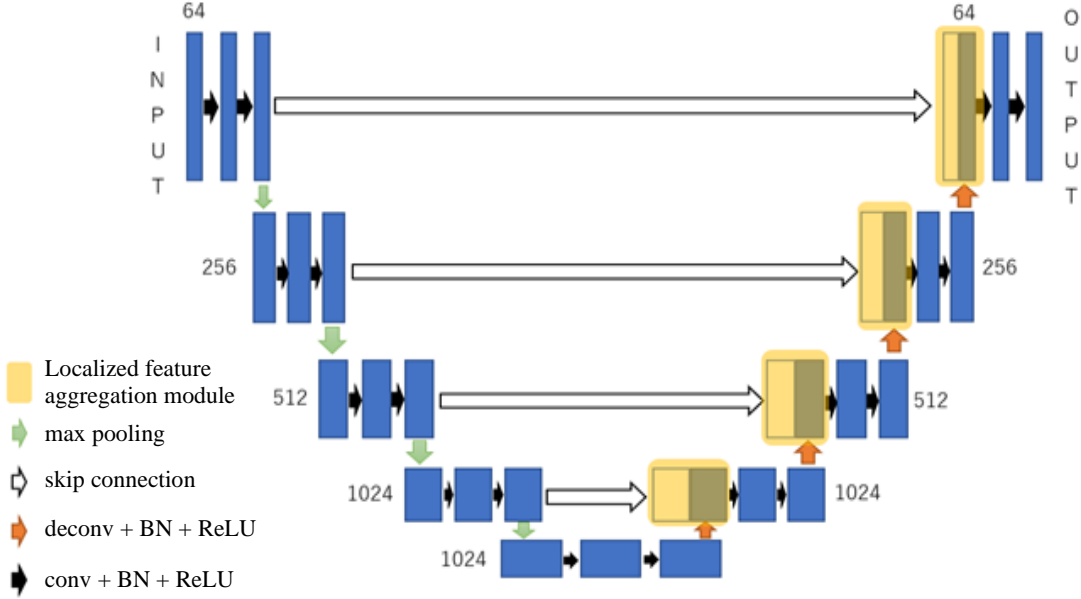

Fig. 2: U-net with localized feature aggregation module

where the height and width of feature map Q are H2 and W2. $Y_p$ emphasizes and aggregates the feature maps of decoder that have high similarity to the feature maps of encoder. Since we aggregate the feature maps of decoder that are similar to the feature maps of encoder which is superior in positional information, we can obtain feature maps that have superior positional information in comparison with the original feature maps of decoder.

The output of feature aggregation module is the sum of $Y_p$ and the feature maps from the encoder by residual connection. We conducted an ablation study on residual connections.

### B. Localized feature aggregation module

From Equation 1, it can be seen that the source-target-attention requires a very large computational cost when the spatial dimension of input feature map is large. The computation of self-attention also requires a very large computational cost under the same conditions. There are many studies for reducing the computational cost of self-attention [6-9]. One of the study is the self-attention method that are restricted to a local range. The computational cost of self-attention becomes $(HW)^2$ where H and W are the size of an input feature map. The computational cost of local self-attention becomes $(m^4 N)$ where m is a local range in Fig.1 and N is number of local ranges. Comparing the computational cost of the two methods, we can see that the local range is more effective when the spatial dimension of the input is large. We also limit the range used to reduce the computational cost of the feature aggregation module. By limiting the range, it is possible to aggregate the information between different features while reducing the computational cost of the feature aggregation module. We defined the local range m as 7 by experiments.

Fig. 2 shows the U-net with the proposed module. The feature maps in the encoder are sent by skip connection and are defined as the Key and Value of localized feature aggregation module. The feature maps in the decoder are defined as the source in the proposed module. The feature maps in the encoder and decoder are aggregated by our module. These processes are done before the convolutional layer in each decoder layer. We conducted ablation studies on whether the encoder or the decoder should be the source or the target. Also, the proposed module was adopted to the feature maps of 1024, 512, 256 and 64 channels with U-net architecture decoders as shown in Fig.2.

## IV. EXPERIMENTS

### A. Dataset and metrics

We used the Drosophila cell image dataset [14]. The dataset consists of 5 classes; membrane, mitochondria, synapses, extracellular and intracellular. Since the original size is 1024×1024 pixels, we cropped a region of 256×256 pixels from original images due to the size of GPU memory. There is no overlap for cropping areas, and the total number of crops is 320. We used 192 regions for training, 64 for validation and 64 for test. We evaluate our method with 5 fold cross-validation.

We also used COVID-19 image dataset [28]. The dataset consists of 4 classes; background, lungs other, ground glass and consolidations. Total number of images is 100. We used 50 for training, 25 for validation and 25 for test. We evaluate our method with 4 fold cross-validation.

In semantic segmentation, Intersection over Union (IoU) is used as evaluation measure. IoU is the overlap ratio between prediction and ground truth labels. In this paper, we used IoU of each class and mean IoU which is the average IoU of all classes.

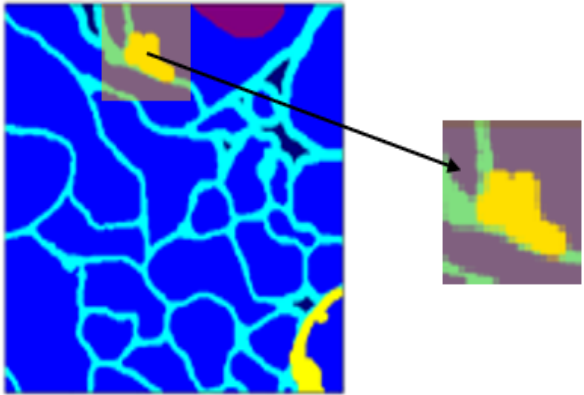

Fig.3: The synapse class with a small number of pixels has a relatively large number of pixels in the local region.

*B. Experimental setup*

In the Drosophila cell image dataset, we used the Pytorch library and trained each network using Adam for 1500 epochs with base learning rate (lr_base) of 0.001. We use a batch size of 8 and a momentum of 0.9. Loss function is the combination of the Focal loss [15] and IoU loss [26] in all methods.

In COVID-19 dataset, we also used the Pytorch library and trained each network using SGD for 250 epochs with base learning rate (lr_base) of 0.001. Learning rate (lr) is scheduled as

$$lr = lr\_base * \frac{\cos\left(\frac{epoch}{max\_epoch}\pi\right)+1}{2} \quad (3)$$

We used a batch size of 8 and a momentum of 0.9. Softmax cross entropy loss was used. Since the class ratio of COVID-19 image dataset is unbalanced, we used class weights [16] to softmax cross entropy loss.

We evaluated the following six methods; U-net, Attention U-net [19], R2U-Net [27] SA-UNet [18], U-net with self-attention, U-net with our method. R2U-Net is a network model that each convolutional layer is replaced by a recurrent convolutional layer. SA-UNet introduces a spatial attention module at the bottom layer of the U-net. In addition, SA-UNet uses structured dropout convolutional blocks instead of the original convolutional blocks of U-Net to prevent network overfitting. U-net with self-attention refers to the model with self-attention at the bottom layer of the U-net.

*B. Experimental results*

We verified whether the computational cost can be reduced by restricting the local range. We evaluated the proposed method when the size of input image is 256x256 pixels. We compared the difference in computational cost when the source-target attention between encoder and decoder is adopted to local range and whole image. By restricting the computation to the local range, total computational cost is 0.0466 times in comparison with the case that the proposed method is adopted to whole image. This result shows that the computational cost of the proposed method is sufficiently reduced.

The accuracy of six methods on the Drosophila cell image dataset is shown in Table 1. Bold letters in the Table indicate the best accuracy. Table 1 shows that our method achieved the best mIoU 72.8% on the Drosophila cell image dataset. We confirmed that the accuracy of our method was 1.46% higher than U-net and 0.83% higher than Attention-U-net. The results show that our feature aggregation between encoder and decoder is superior to conventional concatenation for positional information. The accuracy of RU-Net was 72.69%, SA-UNet was 72.18%, and U-net with self-attention was 72.01%. The accuracy of the proposed method was better than that of the existing methods because the accuracy of synapses class by our method was higher than that of the existing methods.

As shown in Fig. 3, by using the proposed method in a local range, the percentage of the classes with a small number of pixels in the local range increases. Therefore, the proposed method can generate an effective attention map for a class with a small number of pixels, and the accuracy was improved. However, the other classes did not show the same improvement as synapses because the other classes have a relatively large number of pixels.

Fig.4 shows segmentation results on the Drosophila cell image dataset. From left to right, input image, ground true image, the results by U-Net, U-net with our module are shown. We can see that the accuracy of synapse class was improved by using the proposed method. This result shows that the proposed method generates an effective attention map for a class with a small number of pixels. For the other classes, we cannot see any obvious difference in the segmentation results between U-net and our method.

Table 2 shows the accuracy on the COVID-19 image dataset. The proposed method achieved the best accuracy of 46.93% on mean IoU. The proposed method is more accurate than the existing methods on the classes of lungs other and consolidations that class with small number of pixels. By improving the accuracy of difficult classes, mean IoU of our method was about 2% higher than the other methods. From the results on two kinds of image datasets, the effectiveness of our method is demonstrated.

Fig.5 shows segmentation results on the COVID-19 image dataset. From left to right, input image, ground true image, the results by U-Net, U-net with our module are shown. We can see that the misclassification was reduced by the proposed method in comparison with U-net. The results show the effectiveness of the proposed method for recovering positional information.

*C. Ablation study*

We conducted two ablation studies on validation dataset in the COVID-19 validation dataset in order to confirm the effectiveness of the proposed method. In the first ablation study, we evaluated the size of the local range in the localized feature

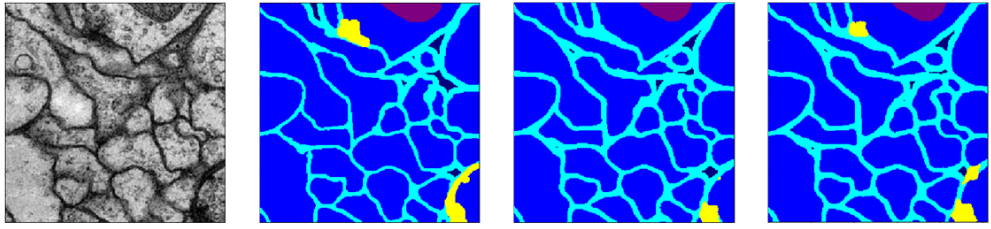

Fig.4: Segmentation results on the Drosophila cell image dataset. From left to right, input image, ground truth, the result by U-net, the result by the proposed method.

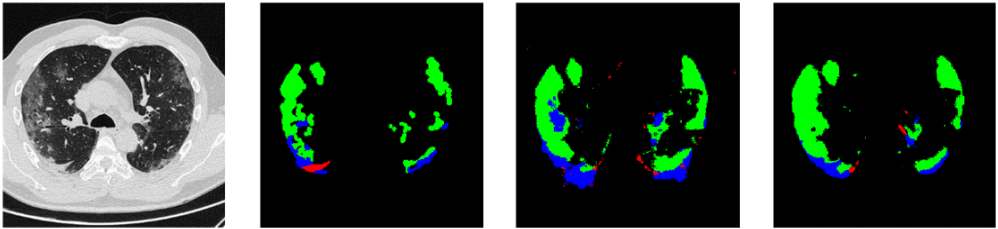

Fig.5: Segmentation results on the COVID-19 dataset. From left to right, input image, ground truth, the result by U-net, the result by the proposed method

aggregation module to determine which size is more accurate with low computational cost. In the second ablation study, we evaluated residual connection of the localized feature aggregation module.

**Local range.** We evaluated the size of the local range in our module. The results are shown in Table 3. The results showed that there is no significant difference in accuracy between the local range sizes of 7 and above. Since the computational cost of localized feature aggregation module increases as the local range increases, we defined the local range of localized feature aggregation module to be 7.

**Residual methods.** We use residual connection in feature aggregation module. We evaluated the accuracy while changing the feature maps for residual connection. The feature maps in encoder, decoder and none are evaluated. The results are shown in Table 4. The results show that it is effective to use the feature maps in encoder for the residual connection. The accuracy improvement can be attributed to the fact that the recovery of positional information is important for semantic segmentation. Thus, it was effective to use encoder's feature maps with superior positional information for residual connection.

## V. CONCLUSION

In this paper, we proposed a new feature aggregation method that computes the similarity between the feature maps in an encoder and a decoder. In addition, we reduced the computational cost by using a locally limited range of attention. We used the proposed module with standard U-Net and confirmed the accuracy improvement on the Drosophila cell image dataset and the COVID-19 image dataset. Experimental results demonstrated that we succeeded in improving the accuracy of difficult class with small number of pixels.

The proposed method improved the accuracy of class with a small number of pixels in comparison with the conventional methods, but the accuracy of other classes was not so improved. The methods using self-attention was improved by introducing positional encoding [6-7]. To improve the accuracy further, we would like to consider how to effectively provide the positional encoding to the proposed method.

REFERENCES

[1] H. Zhao, X. Qi, X. Wang, and J. Jia. "Pyramid Scene Parsing Network." IEEE Conference on Computer Vision and Pattern Recognition. 2017. p. 2881-2890.

[2] L.-C. Chen, G. Papandreou, I. Kokkinos, K. Murphy, and A. L. Yuille. "Deeplab: Semantic image segmentation with deep convolutional nets, atrous convolution, and fully connected crfs." IEEE transactions on pattern analysis and machine intelligence, 2017, p.834-848.

[3] Ronneberger, O., Fischer, P., and Brox, T. "U-net: Convolutional networks for biomedical image segmentation." In International Conference on Medical image computing and computer-assisted intervention, pp.234-241. 2015.

[4] Bahdanau, D., Cho, K., Bengio, Y. "Neural machine translation by jointly learning to align and translate." arXiv preprint arXiv:1409.0473. 2014.

[5] Vaswani, A., Shazeer, N., Parmar, N., Uszkoreit, J., Jones, L., Gomez, A. N., ... and Polosukhin, I. "Attention is all you need." In Advances in neural information processing systems, p. 5998-6008, 2017.

[6] Ramachandran, P., Parmar, N., Vaswani, A., Bello, I., Levskaya, A., and Shlens, J. "Stand-alone self-attention in vision models." arXiv preprint arXiv:1906.05909, 2019.

[7] Wang, H., Zhu, Y., Green, B., Adam, H., Yuille, A., and Chen, L. C. "Axial-deeplab: Stand-alone axial-attention for panoptic segmentation." In European Conference on Computer Vision, pp.108-126. 2020.

[8] Huang, Z., Wang, X., Huang, L., Huang, C., Wei, Y., and Liu, W. "Ccnet: Criss-cross attention for semantic segmentation." I n Proceedings of the IEEE/CVF International Conference on Computer Vision, pp.603-612. 2019.

[9] Ho, J., Kalchbrenner, N., Weissenborn, D., and Salimans, T. "Axial attention in multidimensional transformers." arXiv preprint arXiv:1912.12180. 2019.

[10] H. Zhao, X. Qi, X. Wang, and J. Jia, Pyramid Scene Parsing Network, IEEE Conference on Computer Vision and Pattern Recognition. p. 2881-2890, 2017.

Table 1: Accuracy on the Drosophila cell image dataset.

| Method | Mean IoU(%) | Membrance(%) | Mitcondria(%) | Synapse(%) | Extracellular(%) | Intracellular(%) |
|---|---|---|---|---|---|---|
| U-net | 71.34 | 73.16 | 81.16 | 43.35 | 66.23 | 92.78 |
| Attention U-Net[18] | 71.97 | 72.76 | **82.57** | 45.14 | 66.73 | 92.65 |
| RU-Net(time-step=2)[27] | 72.20 | 73.34 | 82.49 | 45.16 | 67.05 | **92.94** |
| SA-UNet[19] | 72.18 | **73.36** | 82.08 | 45.24 | **67.24** | 92.95 |
| U-net + self-attention | 72.01 | 73.15 | 82.14 | 46.36 | 65.64 | 92.77 |
| U-net with ours | **72.80** | 73.22 | 82.43 | **49.06** | 66.56 | 92.71 |

Table 2: Accuracy on the COVID-19 dataset.

| Method | Mean IoU(%) | Background(%) | Lungs other(%) | Ground glass(%) | Consolidations(%) |
|---|---|---|---|---|---|
| U-net | 44.33 | 95.06 | 31.03 | 43.37 | 7.86 |
| Attention U-Net[18] | 44.43 | 94.91 | 31.07 | 42.90 | 8.86 |
| RU-Net(time-step=2)[27] | 44.97 | 95.62 | 31.29 | 44.85 | 8.12 |
| SA-UNet[19] | 44.75 | **95.73** | 29.38 | **48.88** | 5.02 |
| U-net + self-attention | 44.69 | 95.15 | 31.73 | 42.31 | 9.58 |
| U-net with ours | **46.93** | 95.49 | **36.00** | 44.69 | **11.55** |

Table 1: Comparison results with different local range.

| local range | Method | Mean IoU(%) | Background(%) | Lungs other(%) | Ground glass(%) | Consolidations(%) |
|---|---|---|---|---|---|---|
| | 3 | 46.62 | 94.32 | 36.61 | 40.13 | **15.44** |
| | 5 | 46.77 | 95.25 | 36.09 | 44.00 | 11.74 |
| | 7 | 46.95 | 95.19 | **37.32** | **45.05** | 10.25 |
| | 9 | 46.74 | 95.30 | 35.05 | 44.82 | 11.77 |
| | 11 | **47.05** | **95.37** | 35.87 | 44.97 | 12.00 |

Table 4: Comparison results when the feature maps for residual connection in localized feature aggregation module is changed.

| Method | Mean IoU(%) | Background(%) | Lungs other(%) | Ground glass(%) | Consolidations(%) |
|---|---|---|---|---|---|
| Encoder | **46.95** | **95.19** | **37.32** | **45.05** | **10.25** |
| Decoder | 42.03 | 93.49 | 30.48 | 37.79 | 6.39 |
| None | 41.97 | 94.03 | 28.31 | 35.42 | 10.12 |

[11] L.-C. Chen, G. Papandreou, I. Kokkinos, K. Murphy, and A. L. Yuille, Semantic image segmentation with deep convolutional nets and fully connected crfs. International Conference on Learning Representations 2015.
[12] L.-C. Chen, Y. Zhu, G. Papandreou, F. Schroff, and H. Adam. Encoder-decoder with atrous separable convolution for semantic image segmentation. In Proceedings of the European conference on computer vision, p. 801-818. 2018.
[13] Wu, Huikai, et al. "Fastfcn: Rethinking dilated convolution in the backbone for semantic segmentation." arXiv preprint arXiv:1903.11816, 2019.
[14] Segmented anisotropic ssTEM dataset of neural tissue. Stephan Gerhard, Jan Funke, Julien Martel, Albert Cardona, Richard Fetter. figshare. Retrieved 16:09, (GMT), 2013.
[15] Lin, T. Y., Goyal, P., Girshick, R., He, K., and Dollár, P. "Focal loss for dense object detection." In Proceedings of the IEEE international conference on computer vision, pp.2980-2988, 2017.
[16] De Boer, Pieter-Tjerk, et al. "A tutorial on the cross-entropy method." Annals of operations research 134.1, pp.19-67, 2005.
[17] Badrinarayanan, V., Kendall, A., and Cipolla, R. "SegNet: A Deep Convolutional Encoder-Decoder Architecture for Image Segmentation" IEEE Transactions on Pattern Analysis and Machine Intelligence. Vol.39, pp. 2481-2495, 2017.
[18] Oktay, O., Schlemper, J., Folgoc, L. L., Lee, M., Heinrich, M., Misawa, K., and Rueckert, D. "Attention u-net: Learning where to look for the pancreas." arXiv preprint arXiv:1804.03999. 2018.
[19] Guo, C., Szemenyei, M., Yi, Y., Wang, W., Chen, B., and Fan, C. "SA-UNet: Spatial Attention U-Net for Retinal Vessel Segmentation." arXiv preprint arXiv:2004.03696. 2020.
[20] Anderson, P., He, X., Buehler, C., Teney, D., Johnson, M., Gould, S., Zhang, L.: Bottom-up and top-down attention for image captioning and vqa. arXiv preprint arXiv:1707.07998, 2017.
[21] Wang, F., Jiang, M., Qian, C., Yang, S., Li, C., Zhang, H., Wang, X., Tang, X.: Residual attention network for image classification. In: IEEE CVPR. pp. 3156–3164, 2017.
[22] Hu, J., Li S., and Gang S. "Squeeze-and-excitation networks." Proceedings of the IEEE conference on computer vision and pattern recognition. pp.7132-7141, 2018.
[23] Fu, Jun, et al. "Dual attention network for scene segmentation." Proceedings of the IEEE/CVF Conference on Computer Vision and Pattern Recognition. pp. 3146-3154, 2019.
[24] Chen, L. C., Yang, Y., Wang, J., Xu, W., and Yuille, A. L. "Attention to scale: Scale-aware semantic image segmentation." In Proceedings of the IEEE conference on computer vision and pattern recognition, pp. 3640-3649. 2016.
[25] Li, Yanwei, et al. "Attention-guided unified network for panoptic segmentation." Proceedings of the IEEE/CVF Conference on Computer Vision and Pattern Recognition, pp.7026-7035, 2019.
[26] Zhou, D., Fang, J., Song, X., Guan, C., Yin, J., Dai, Y., and Yang, R. "Iou loss for 2d/3d object detection." In 2019 International Conference on 3D Vision, pp. 85-94. 2019.
[27] Alom, M. Z., Hasan, M., Yakopcic, C., Taha, T. M., and Asari, V. K. "Recurrent residual convolutional neural network based on u-net (r2u-net) for medical image segmentation." arXiv preprint arXiv:1802.06955. 2018.
[28] Covid-19 CT segmentation dataset. https://medicalsegmentation.com/covid19/ , 2020.
[29] Hu, J., Shen, L., and Sun, G. "Squeeze-and-excitation networks." In Proceedings of the IEEE conference on computer vision and pattern recognition, pp. 7132-7141, 2018.
[30] Wang, X., Girshick, R., Gupta, A., and He, K. "Non-local neural networks." In Proceedings of the IEEE conference on computer vision and pattern recognition, pp. 7794-7803, 2018.
[31] Cao, Yue, et al. "Gcnet: Non-local networks meet squeeze-excitation networks and beyond." Proceedings of the IEEE/CVF International Conference on Computer Vision Workshops, pp. 0-0, 2019.
[32] Liu, Jiang-Jiang, et al. "Improving convolutional networks with self-calibrated convolutions." Proceedings of the IEEE/CVF Conference on Computer Vision and Pattern Recognition, pp. 10096-10105, 2020.
[33] Roy, A. G., Navab, N., and Wachinger, C. "Concurrent spatial and channel 'squeeze & excitation' in fully convolutional networks." In International conference on medical image computing and computer-assisted intervention, pp. 421-429, 2018.
[34] Woo, S., Park, J., Lee, J. Y., and Kweon, I. S. "Cbam: Convolutional block attention module." In Proceedings of the European conference on computer vision, pp. 3-19, 2018.